# THE EFFECT OF USING FACEBOOK MARKUP LANGUAGE (FBML) FOR DESIGNING AN E-LEARNING MODEL IN HIGHER EDUCATION

Mohammed Amasha[1], Salem Alkhalaf[2]

[1]Faculty of Specific education, Computer Department, Domyat University, EGYPT
Computer Science Department, Qassim University, Saudi Arabia
Email: mw_amasha@yahoo.com

2Faculty of Science and Arts, Computer Science Department, Qassim University, Alrass City, SAUDI ARABIA
Email: s.alkhalaf@qu.edu.sa

*Abstract*: This study examines the use of Facebook Markup Language (FBML) to design an e-learning model to facilitate teaching and learning in an academic setting. The qualitative research study presents a case study on how, Facebook is used to support collaborative activities in higher education. We used FBML to design an e-learning model called processes for e-learning resources in the Specialist Learning Resources Diploma (SLRD) program. Two groups drawn from the SLRD program were used; First were the participants in the treatment group and second in the control group. Statistical analysis in the form of a t-test was used to compare the dependent variables between the two groups. The findings show a difference in the mean score between the pre-test and the post-test for the treatment group (achievement, the skill, trends). Our findings suggest that the use of FBML can support collaborative knowledge creation and improved the academic achievement of participatns. The findings are expected to provide insights into promoting the use of Facebook in a learning management system (LMS).

## I. INTRODUCTION

The challenge of using Web 2.0 technologies in the classroom is to use them in a way that enhances learning, not simply because they are available [4]. Web 2.0 tools offer methods for personalizing classes and demonstrating instructional presence. Some of the more widely recognized tools include blogs, wiki, RSS feeds, video and photo sharing, avatars, microblogging, social bookmarking, and social media [23]. Social media technology and social networking communities (SNCs) have become an essential part of personal life as users generate content [1]. The ubiquity of social media is no more apparent than at universities where the technology is transforming the ways students communicate, collaborate, and learn [22]. Facebook has become the most popular social networking site worldwide. People increasingly use Facebook for communication, social networking, and interaction. Using Facebook becomes part of a routine [12]. Today, Facebook is increasingly important for activism and education. Facebook can be used by professors to follow students' learning process and students themselves [10]. Facebook is a social networking website that supports many features.

The use of collaborative technologies such as Facebook and Twitter leads to an instant online community in which people communicate rapidly and conveniently with each other. However, interest and concern regarding the topological structure of these new online social networks are growing. The existence of such online applications and services as Facebook and YouTube are well known among teachers, who often use this technology themselves in their private lives, but may not recognize the educational potential for their students [8].

Facebook was chosen as the host site for this research because of the uniform strength of its features compared to other popular social networking sites such as MySpace and Friendster, the image-sharing site Flickr, and the open source learning management system Moodle. In this study, we provide a model for learning and teaching using Facebook markup language (FBML), an extension of HTML that works within Facebook Canvas pages [26]. It is based on objective scientific and higher efficiency and productivity by presenting information and skills to students in line with recent trends. This idea is based on learners and on their needs, aspirations, and convictions. The research tried to benefit from modern technology available in computer technology and its impact on technology education.

### A. Research problem

The author(s) observed that students face problems in performing practical tasks, difficulties in exchanging their experience and sources. Also there is a lack of continuous & effective communication between the students and teachers. These problems lead to limited feedback, which was confirmed by a survey of students in the Specialist Learning Resources Diploma (SLRD) program at Qassim University in Saudi Arabia, who participated in the DSL 1107 course. The results showed that the students lacked a desire to learn, motivation, and enthusiasm to perform educational tasks. The results also showed low academic achievement, which affected the educational process.

In this study, we developed a model to be used in FBML to build an e-learning system on the web and then applied the model in a sample to determine it's effectiveness to increase learning and enriching knowledge. The study also examined the views of participants on using Facebook apps in the learning process.

The following questions guided the research:

- What is a suggested model of e-learning content based on Facebook apps using FBML?
- What is the impact of using this model in teaching and managing the course?
- What is the effectiveness of the model in mastering the cognitive aspects and skills by the participants?
- What do the students think about the model?

### B. Research significance

This study examined the effectiveness of the model in students' academic achievement. The current research signifies that we designed a model for e-learning content based on Facebook apps using FBML, measured the effectiveness of the model and its influence on students' academic achievement Then, created a template for an e-learning course to make best use of the teacher and learner's effort, and linked the technological changes with the educational process to meet the challenges of the next century.

### C. Research hypotheses

The following hypotheses guided our work. (1) Using (FBML) is effective in providing e-content to the research sample. (2) There is statistical significance in the mean of students' cognitive achievement after they took the course. (3) There is statistical significance in the mean of the students' skill performance after they took the course. (4) Students have positive views on teaching and the learning method used in this research, and there is a statistical significance in students' trends in using Facebook after they took the course.

## II. LITERATURE REVIEW

### A. What is Facebook?

Facebook is an online social networking service and is a prominent example of an SNS. The large number of Facebook users was cited previously, and among higher education students, its use appears to be even more widespread [21]. It operates somewhat like a personal website, but within a defined community of users and with functions that allow users to locate and interact with each other [28]. Facebook has become a popular social networking possibility for reaching students [21]. It not only allows members to upload photos and change personal profiles online but also allows them to leave messages for friends who are offline. Facebook helps maintain relationships with others when users are not active online [17].

Facebook can be used as a communication method, an instructional resource, a collaborative tool, and a showcase for student projects. Facebook can be used to post class announcements for parents or provide schedule reminders for students. Facebook can be used as an instructional resource, in which teachers can post tips, explanations, or samples to help students learn. Facebook is an example of a Web 2.0 social networking site, which has enormous potential in the field of education although the site was not designed as an environment for constructing and managing learning experiences. Facebook operates as an open platform, unlike other systems organized around courses or formally structured content [2].

In addition to these functions that are common to most traditional social networking sites, Facebook offers a variety of additional functions, such as a news feed, through which users can follow the movement of their Facebook friends who are also users of the system [24].

On Facebook, a teacher can start a closed group that is unavailable in Facebook search and use the group to publish information on which students must provide feedback [18]. In addition, tutors may post homework, links for further study, and announcements, share thoughts, and communicate socially after school hours [16]. Facebook plays a significant role in students' lives and is one of the most resourceful methods for interaction. It is a cumbersome learning process with only one direction of communication.

In a recent interview, instructional technologist Jim Groom pointed out that the there is a problem with the conventional LMS. Such strong opposition deserves scrutiny because Groom has vast experience in the subject matter. The one-way communication (one-sided communication) is either between the student and the instructor or vice versa. Therefore, the course localized the class discussions into discussion boards, and exchange of information among group members becomes slow and burdensome. In the same system, only the instructor has the mandate to review and grade projects submitted by students [5]. Furthermore, Groom thinks that the Facebook LMS creates productive class interaction and links it to the entire world. It puts students in a versatile environment where they feel free to share and learn easily. However, Twitter increases the versatility and flexibility of conversation. It gives a sense and a platform for students to learn. Twitter is becoming an increasingly important tool for students to express their abilities thoughtfully [12].

*B. The benefits of using Facebook*

Facebook is free and is one of the best media for communication. With the help of Facebook, you can connect to different people anywhere in the world, because almost everyone familiar with the Internet uses Facebook. This gives us the opportunity to know more about others' culture, values, customs, and traditions. Facebook is the most appropriate tool for finding old friends. When friends move away, we often do not have the opportunity to communicate with them. However, now Facebook gives us the opportunity to communicate with old friends very easily at no cost. We can share our feelings about what is happening in our daily lives through Facebook. You can also get feedback from your friends about their reactions to our feelings. It is the best medium for sharing one's feelings and thoughts with others. Facebook has good privacy settings, which gives the option of maintaining privacy according to one's preferences. You can chat with friends by using Facebook. The Facebook messenger application is very handy for chatting online. Students can use Facebook for group study by creating a group only for studying. There you can share information about your projects, homework, assignments, exams, due dates, etc. You can also use Facebook as a social bookmarking site. You can share articles, blogs, photos, etc., with thousands of people. You can use Facebook groups to connect all your close friends together. You can also maintain the privacy of the group by making it private. Facebook also has new features such as group chats, notifications, file sharing, etc. These features help members of the group stay connected. Facebook has helped many students become more independent, understanding with the teacher and more academic [15].

Educational Facebook was thought to be an eligible subject for the study because of the following reasons: The number of social network users in intense communication was high, every user knows the settings, and Facebook offers e-mail, forums, and chats similar to a learning management system. It was critical to reveal the reasons for the influence of educational use [9].

*C. Facebook features*

Facebook is very effective for use as an LMS for e-learning. Facebook is used as a tool in enhancing e-learning courses [16]. The latest version of eFront is made with a set of social tools that enable the use of Facebook as an LMS for e-learning. It comes with a simplified Facebook integration plug-in for easy use. EFront is a simplified open source learning management system whose appearance is attractive and is SCORM certified. EFront facilitates community learning and upholds principles of collective knowledge [5]. Using technologies that students are comfortable with such as Facebook help develop a successful learning environment by putting into practice the creative, interactive, and collaborative nature of Facebook. These technologies aim at meeting learners where they are and are useful in helping them achieve their educational purposes [25].

Facebook is among the fastest-growing social networks that assist people in efficient communication with allies, relatives, workmates, and schoolmates. Research shows that about 50% of active Facebook users log on to their Facebook accounts every day [25]. An average Facebook user is linked to about 80 society groups, events, and groups. Facebook can be used as an LMS for e-learning. Most Facebook users are learners and are familiar with effective use of Facebook. Use of Facebook for e-learning could instigate the majority of students to undergo effective learning [5]. Although Facebook was not developed for learning purposes, more than 20 Facebook educational applications exist that involve distinct types of interactions [25].

The three major types of interactions include learner-to-instructor interaction, learner-to-learner interaction, and learner-to-subject interaction. Learner-to-instructor interaction involves contact between the learner and a tutor, professional, or an instructor. The second type of interaction involves electronic dialogue, online classroom debates, and print or electronic dialogue [5]. Learner-to-learner interaction may happen in a group situation or outside a group. The third type of interaction involves discussions conducted by learners and controlled by the teacher. Alternatively, learners may be allowed to act independently in group projects or group-head activities. Learner-to-content interaction involves interaction between learners and the topic at hand. It is believed that interaction between a learner and the subject topic [18]. Content Slide Share is used to generate and share presentations using Facebook. Also, it is possible to add documents, PDFs, and MP3 to develop webinars. Instructors can distribute mathematical formulas and workouts to learners [3]. The good thing about this application is that learners can view the formulas even if the learners do not have the application themselves. News Rack is a completely featured RSS reader used as an impartial client with Google Readers; other information relevant to the course can be shared with learners via this application [16]. It is possible to create an event calendar, amend and keep a monthly calendar in the profile, and share it with others learners. Finally, it is possible to create flashcards to assist studying on Facebook.

According to Warburton [25], Learners Neat ChatLearners allows chatting on Facebook pages and groups. Learners share files, send confidential messages, and perform much more intuitively and faster. Study groups allow easy collaboration with fellow learners [3]. Tutor can make homework arrangements for the course. Books iReadShare is an application that lets classmates share books. Book tag is an application used by learners to create arbitrary booklists [16]. Learners label books and share them with fellow learners. PeerPong is an application that allows learners to discover and connect them with experts who can answer their queries professionally [25].

According to Mao [16], instructor Survey Gimso is an application that learners use to create and share surveys, polls, and short assessments on Facebook. UdutuTeach is an application that allows importation of courses from my Udutu (authoring tool). It is used in managing the type of learners who can take a certain course and track learners' progress. Webinaria Screen-cast Recorder is used by tutors to create screen-cast lectures. The screenshots can then be shared with the learners on Facebook. Quiz monster is a Facebook quiz application [16].

*D. Other studies that examined the same issues*

The purpose of the Guzin and Kocak [7] study was to design a structural model that explained how users could utilize Facebook for educational purposes. To shed light on the educational use of Facebook, in constructing the model, the relationship between users' Facebook adoption processes and their educational use of Facebook was included indirectly while the relationship between users' purposes in using Facebook and the educational usage of Facebook was included directly.

The goal of Shih's [19] study was to investigate the effect of integrating Facebook and peer assessment with college English writing class instruction through a blended teaching approach. This blended approach consisted of one-third of a semester of classroom instruction and two-thirds of a semester combining Facebook, peer assessment, and classroom instruction. The subjects were 23 first-year students majoring in English at a technological university in Taiwan who participated in an 18-week English writing class. The students were divided into three groups with three Facebook platforms. Quantitative and qualitative approaches were used in the study. Research instruments included a pre-test and post-test of English writing skills, a self-developed survey questionnaire, and in-depth student interviews. The findings suggest that incorporating peer assessment using Facebook in learning English writing can be interesting and effective for college-level English writing classes. Students can improve their English writing skills and knowledge with not only the in-class instruction but also cooperative learning. In addition, Facebook-integrated instruction can significantly enhance students' interest and motivation. Finally, the findings may provide useful instructional strategies for teachers of English as Second Language (ESL) English writing courses.

Lenandlar [11] explored the use of Facebook groups in the undergraduate computer science program at the University of Guyana. Specifically, guided assessment strategies using Facebook groups were compared with unguided and non-assessed Facebook groups. This study provides a comparative outline of the usage patterns of two instructor-guided and assessed Facebook groups with three student-led, no assessed Facebook groups that supported a form of open discourse. Results suggest that planned and guided, instructor-directed activities provide more focused responses from students compared to open discourse.

Yamauchi, Fujimoto, Takahashi, Araki, Otsuji, and Suzuki [27] examined the Socla study program to build a social learning community for high school students using Facebook and other Internet services. In the two-week program, the students worked on individual study projects that focused on the students' future plans. With the help of volunteer supporters and facilitators, the students found relevant information and received constructive feedback about their progress. The results indicated that the students viewed their own future more positively, realized that learning about unknown subjects can be interesting, and discovered that advice from their superiors was useful. Responses from the volunteer supporters showed that the program also worked as a reflective learning opportunity. Moreover, the authors identified several issues through the program to address for the success of this approach.

According to Lovell and Palmer's [14] study, undergraduates' use of social networking sites has been well documented in the popular press and academic publications. Research suggests that students spend, on average, 30 minutes a day engaged in a predictable routine of social networking. Correspondingly, on the first author's previous campus, she had frequently observed many of the students in her introductory writing class spending time in the Highlands College library enthusiastically communicating on Facebook. In this study, the researchers harnessed Highlands College students' enthusiasm for social networking to increase their feelings of connectedness and improve critical thinking and writing skills through Facebook assignments.

## III. METHODS

### A. Research methods

We used the experimental method, since it involves study factors and variables that affect the phenomenon or problem and changed several aspects with other fixed variable to reach the causative relationship between these variables.

### B. Participants

The participants included 58 students from the SLRD program at Qassim University in Saudi Arabia. The students were randomly divided into two groups. The first group (experimental) consisted of 29 students, and the second group (the control) consisted of 29 students.

### C. Case study

Designing and teaching a course that involved the use of a social network was chosen as an online platform to create a class on Facebook. Students created their own profiles, uploaded photos, audio, podcasts, and video, created and joined discussion groups, sent messages, published blogs, and made presentations. A group was created on a Facebook page (processes- for- eLearning resources). Students were invited to join the group (29 students). A page was created on Facebook by using FBML (http://statictab.com/r4g72y). Twelve lectures were uploaded to the group page. A chat panel was added to receive questions and inquiries of the students. Teamviewer software was used to follow the work of the students online and edit it. Google Drive was used to create e-tests and add them to the page. If a media file or text was uploaded on the Facebook page, the Twitter feed was used to send notifications to students on their Twitter accounts. Only class members in the course were invited to join the class on the Facebook site. No guest or outsider was allowed to join or participate.

### D. Selected content

The "Processes for E-learning Resources" course was chosen as the educational content for the Facebook site, and then the content was uploaded on the website as files: videos, texts, images, and flash files. Then the content was divided into lectures. We allowed students to comment and add their viewpoints on all items.

### E. Validity of the content

To ensure the veracity of the content, we presented the research tools to experts to express their opinions on the content. Some points were corrected. Twelve students were selected to examine the validity and reliability of the research tools.

### F. Achievement test

The aim of the test was to measure the students' achievement in knowledge and information in the educational content. The Google Drive application was used to design the test. Fifty test questions included different types, such as multiple-choice, true/false, and fill in the blank.

*G. Skills test*

The aim of the designed test was to measure how well the students' learned during the course.

*H. Descriptive measure and statistical analysis*

The questionnaire was designed in order to identify the effectiveness of the use of Facebook in teaching the course. The questionnaire consisted of 20 phrases and the presented to a group of experts to ensure its validity and reliability. To statistically test the validity of the phrases, a correlation coefficient was calculated for each phrase by the total degree of the dimension to which it belong, void of phrase degree. The findings then indicated that the terms of the measure were characterized by a feasible degree of validity. The validity of the measure was calculated with the statistical program LISREL. SPSS software was used to analyze the final results.

## IV. DISCUSSION

*A. E-content*

Using Facebook was effective in providing e-content to the research sample. To test the validity of this assumption, Table 1 shows the statistical significance of the means of the degree of the two groups using a t-test. Results indicate that there is a statistically significant distinction between the mean of the students' degrees of the first group (the pre-group) in favor of the second group (the post-group). The distinction is essential t(28, N = 29) = 30.451; p = .05.

*Table 1: t-test for the statistical significance of means of the degree of the two groups*

| Group | M | sd | df | t | Sig. |
|---|---|---|---|---|---|
| Pre-group | 15.52 | 4.306 | 28 | 30.451 | 0.000 |
| Post-group | 45.41 | 3.96 | | | |

$^*$P <.05. **p<.01.

*B. Students' cognitive attainment*

There was statistical significance between the means of students' cognitive attainment after the course. Table 2 shows the statistical significance of the means of the degree of the two groups using a t-test. Results indicate that there is a statistically significant distinction between the mean of the students' degrees of the experimental group in favor of the control in the students' cognitive attainment after teaching for knowledge, t(56, N = 29) = 21.173; p = .05.

*Table 2: t-test for the statistical significance of means of the degree of the two groups*

| Group | M | Sd | df | t | Sig. |
|---|---|---|---|---|---|
| Experimental | 45.41 | 3.96 | 56 | 21.173 | 0.000 |
| control | 17.79 | 5.79 | | | |

$^*$P <.05. **p<.01.

*C. Students' skill performance*

There was statistical significance between the mean of the students' skills performance after the course. Table 3 shows the statistical significance of the means of the degree of the two groups using a t-test. Results indicate that there is a statistic significant distinction between means of the students' degrees of the experimental group in favor of the control group in students' cognitive attainment after teaching for skills, t(56, N = 29) = 42.616; p = .05.

*Table 3: t-test for the statistical significance of means of the degree of the two groups*

| Group | M | Sd | df | t | Sig. |
|---|---|---|---|---|---|
| Experimental | 21.14 | 7.958 | 56 | 42.616 | 0.000 |
| Control | 92.72 | 4.300 | | | |

$^*$P <.05 **p<.01

*D. Use of Facebook*

Students had positive views of the teaching and learning method used in this research, and there was statistical significance between students' use of Facebook. To test the validity of this assumption, the researchers used the general method of calculating the $\chi^2$ of the frequency table $1\times 2$ to work out the statistical significance of the frequency distinctions between the students' approval and disapproval of each phrase of the questionnaire the using class Facebook site to teach and learn e-courses. Table 4 shows the $\chi^2$ values for students' attitudes toward using a Facebook page as a learning method.

*Table 4: Shows f, p and $X^2$ values, for student' attitudes towards using using Facebook page as a learning method.*

| N | Items | Agree | | disagree | | $X^2$ |
|---|---|---|---|---|---|---|
| | | F | % | f | % | |
| 1 | I feel comfort to use the class Facebook page | 22 | 75.9 | 7 | 24.1 | 7.76 |
| 2 | Facebook page allows me to interact and build a learning community. | 23 | 79.3 | 6 | 20.7 | 9.97 |
| 3 | Facebook page allows me to personalize page to express individuality and creativity. | 25 | 86.2 | 4 | 13.8 | 15.2 |
| 4 | Facebook page allows me to share photos, music and videos. | 25 | 86.2 | 4 | 13.8 | 15.2 |
| 5 | Facebook page allows me to pose questions to the community. | 23 | 79.3 | 6 | 20.7 | 9.97 |
| 6 | Facebook page allows me to hold forums to discuss topics interact. | 28 | 96.6 | 1 | 3.4 | 25.1 |
| 7 | Facebook page allows me to find and share educational resource. | 24 | 82.8 | 5 | 17.2 | 12.4 |
| 8 | Facebook page allows me to create a study group. | 25 | 86.2 | 4 | 13.8 | 15.2 |
| 9 | Facebook page encourages learners centered activities. | 22 | 75.9 | 7 | 24.1 | 7.76 |
| 10 | Facebook page provides collaborative learning opportunities. | 25 | 86.2 | 4 | 13.8 | 15.2 |
| 11 | Facebook page allows me to share Facebook, myspace, digg, delicious and twitter. | 26 | 89.7 | 3 | 10.3 | 18.2 |
| 12 | Facebook page gives me sense of belonging. | 25 | 86.2 | 4 | 13.8 | 15.2 |
| 13 | Facebook page promotes knowledge sharing. | 23 | 79.3 | 6 | 20.7 | 9.97 |
| 14 | Facebook page is a friend of the user. | 29 | 100 | 0 | 0 | 29.00 |
| 15 | Facebook page allows me to send and receive information from other members through messages. | 26 | 89.7 | 3 | 10.3 | 18.2 |
| 16 | Facebook page is a great tool for class communication. | 26 | 89.7 | 3 | 10.3 | 18.24 |
| 17 | I will become actively involved in courses that use Facebook page | 27 | 93.1 | 2 | 6.9 | 21.55 |
| 18 | Facebook page can be used for professional development. | 22 | 75.9 | 7 | 24.1 | 7.76 |
| 19 | I would like to see more Facebook page use in other class. | 27 | 93.1 | 2 | 6.9 | 21.55 |

*P <.05. **p<.01.

Table 4 shows that the $\chi^2$ value of all phrases concerned with the students' viewpoints on using a class Facebook site is bigger than the $\chi^2$ (1, N = 29) = 3.84; p =.05. This indicates that the distinction between the frequencies (f) and the expected frequency (fe) for the phrases concerned were statistically significant. The distinction between the students' approval or disapproval of the phrases of the questionnaire was in favor of the use of using Facebook site in learning and teaching method.

In other words, the students felt comfortable using the class Facebook site. It allowed them to personalize their pages to express individuality and creativity; share photos, music and videos; send and receive information from other members through messages; and hold forums to discuss topics interact. They also described the Facebook site as a great tool for class communication. In addition, there was general satisfaction among the students because they sued Facebook applications in other classes. The students clarified that using the class Facebook site was better than traditional learning methods. The students, furthermore, strongly emphasized the importance of using class Facebook sites in all classes.

The results indicate the existence of statistical difference between the means of students' attitudes toward the use of Facebook in learning and teaching. Table 5 shows the t-test for the statistical significance of the means of the degree of the two groups. The table shows the existence of a statistically significant difference between the means of the two groups (pre-post) toward the use of Facebook site in teaching and learning for students after learning.

*Table 5: t-test for the statistical significance of means of the degree of the two groups.*

| Group | M | Sd | t-test | df | Sig. |
|---|---|---|---|---|---|
| Pre-group | 35.31 | 2.64 | 29.034 | 28 | 0.000 |
| Post-group | 20.10 | 1.47 | | | |

[*]$P < .05$ [**]$p < .01$.

## V. CONCLUSION

Using FBML in Facebook Canvas can help to develop an e-learning course with social media and supports collaborative knowledge creation and sharing in the academic environment.

The use of Facebook is effective in learning and teaching. There was statistical significance between the means of students' cognitive attainment after the course and after teaching for skills performance. Students were positive about the teaching and learning method used in this research, and there was a statistical significance between the means of the student's attitudes toward using Facebook in the classroom. Web 2.0 technology and its components can be used to develop a good learning and teaching system at the undergraduate level.

## VI. RECOMMENDATIONS

We recommend the following practices: Take advantage of Web 2.0 technology in designing e-learning courses, train students to use SNS applications in learning, link SNS applications with the LMS used in universities, and support research e-learning and distance education.